\documentclass[a4paper,11pt]{article}
\usepackage{jheppub} 

\newcommand{\n}{\nonumber}
\newcommand{\p}{\prime}

\usepackage{jheppub}
\usepackage{color}
\usepackage[dvipsnames]{xcolor}
\usepackage{graphicx}
\usepackage{wrapfig, enumerate, slashed}

\usepackage[noabbrev]{cleveref}

\usepackage{footmisc}
\usepackage{amsmath}
\usepackage{wasysym} 
\usepackage{graphicx}
\usepackage{color}
\usepackage{comment}
\usepackage{hyperref}

\newcommand{\bee}{\begin{eqnarray}}
\newcommand{\eee}{\end{eqnarray}}
\newcommand{\beeq}{\begin{equation}}
\newcommand{\eeeq}{\end{equation}}

\newcommand{\beqa}{\begin{eqnarray}}
\newcommand{\eeqa}{\end{eqnarray}}
\newcommand{\be}{\begin{equation}}
\newcommand{\ee}{\end{equation}}
\newcommand{\ba}{\begin{array}} 
\newcommand{\ea}{\end{array}}
\def\bea{\begin{eqnarray}}
\def\eea{\end{eqnarray}}
\newcommand{\nn}{\nonumber}

\newcommand{\Dfb}{\mathord{\buildrel{\lower3pt\hbox{$\scriptscriptstyle{\leftrightarrow \tiny{ \ \ \ } }$}}\over {D^{\mu}}}}

\newcommand{\Dfbd}{\mathord{\buildrel{\lower3pt\hbox{$\scriptscriptstyle\leftrightarrow$}}\over {D}_{\mu}}}

\usepackage{bbold}
\usepackage{comment}

\begin{document}
\title{Gauge Invariant Effective Potential}
\author{Debanjan Balui, Joydeep Chakrabortty,  Debmalya Dey, Subhendra Mohanty}
\affiliation{Indian Institute of Technology Kanpur, Kalyanpur, Kanpur 208016, Uttar Pradesh, India}
\emailAdd{debanjanb22}
\emailAdd{joydeep}
\emailAdd{debmalyad23}
\emailAdd{mohantys@iitk.ac.in}

\abstract{We show that the long-standing problem of gauge dependence of the effective potential arises due to the factorisation of the determinant of operators, which is invalid when we take the zeta-regularised trace of the operators. We show by correcting for this assumption by computing the multiplicative anomaly, the gauge-dependent terms of the effective potential cancel. We also show that in two- and odd-dimensional non-compact spacetime manifolds where the multiplicative anomaly term is zero, the standard calculation of one-loop effective potential gives a gauge-independent result. These results are in support of our claim that the multiplicative anomaly may play a crucial role in removing the gauge dependence in the effective potential in the four-dimensional non-compact manifold. Noting the non-trivial aspects of this anomaly computation for a generic scenario, we propose the Heat-Kernel method to compute the effective potential where this anomaly emerges as a total derivative, thus redundant. We explicitly show how one can calculate the gauge independent, effective action and the Coleman-Weinberg effective potential by employing the Heat-Kernel method. Based on this result, we advocate the Heat-Kernel expansion as the most straightforward method, as it naturally deals with the matrix elliptic operator for the calculation of manifestly gauge independent, effective actions compared to other conventional methods.  }

\maketitle

\section{Introduction}
Effective theories are useful tools for capturing the dynamics of UV complete theories at energies where the heavy masses of the UV theories are inaccessible. By integrating out the heavy scalars or fermions from the UV theory, the effective Lagrangian is expressed as a series of higher dimensional operators suppressed by powers of the integrated out particle mass $1/m^2$. Another context in which effective theories are used is in splitting the fields into ultra-violet and infrared modes with respect to some selected scale and integrating out the ultraviolet modes to obtain classical dynamics of the theory in a series of $g^2 \hbar$ loop corrections to the classical Lagrangian. An example of this is the Coleman-Weinberg (CW) potential \cite{Coleman:1973jx} of the Higgs, which is used for studying the quantum corrected classical dynamics of the Higgs field like symmetry breaking and phase transitions in the early universe \cite{Sher:1988mj}.  

\noindent
One long-standing problem in the calculation of effective theories, first pointed out by Jackiw \cite{Jackiw:1974cv},  is that the effective potential of a gauged scalar is gauge-dependent. The gauge dependence persists whether the effective potential is calculated using the functional method \cite{Dolan:1974gu} or the perturbative diagram method \cite{Weinberg:1973ua}. A way to use the effective potential to calculate physically meaningful quantities was pointed out by Nielsen \cite{Nielsen:1975fs}, who showed that the minima and maxima were independent of gauge choice. This gauge-independent extremum implies that in the broken symmetric minima, the mass spectrum of particles is gauge-independent \cite{Aitchison:1983ns}. Thus, for applications of phase transitions at finite temperature in the early universe, quantities like the critical temperature $T_c$, which are defined using the potential at the values of the field at the true and false vacuum minima, are gauge independent \cite{Metaxas:1995ab, Patel:2011th}. One application where the gauge dependence of the CW potential is a problem is the application of the inflaton potential. Here, for comparison with observations, we need the potential and its derivatives in the slow-roll phase (and not at the extrema) \cite{Urbano:2019ohp}. Various aspects of this problem have been studied over the years \cite{Tye:1996au, DiLuzio:2014bua, Andreassen:2014eha, Andreassen:2014gha, Martin:2014bca, Espinosa:2016uaw, Irges:2017ztc, Chigusa:2018uuj, Chigusa:2017dux, Shen:2021tqk} and the consensus is that the one loop effective action can be gauge dependent while physical quantities which only depend on the extrema of the potential are gauge invariant.

 We organise the paper as follows. In section \ref{sec:JD-pot}, we summarise the effective potential computation of Dolan-Jackiw \cite{Dolan:1974gu, Jackiw:1974cv} in case of a massless scalar QED theory at $d=4$. There we identify the well-known gauge parameter dependency in the one loop effective potential. Then, in section \ref{sec:zeta-function-ma}, we discuss the emergence of multiplicative anomaly associated with Zeta-function regularisation when one loop functional determinant is decomposed in terms of a product of determinants of elliptic operators. We show how to estimate this anomaly in case of non-compact even dimensional manifold. This anomaly vanishes for two- and odd-dimensional non-compact manifolds. We show how the inclusion of this anomaly in the effective action exactly cancel the gauge parameter dependent term and provide a gauge independent effective potential.  We confirm our claim by computing the effective terms in the potential for the similar massless scalar QED theory in $d=2$, and $3$. In the subsequent section \ref{sec:HK-Veff}, we introduce the Heat-Kernel method to compute the effective potential where unlike the conventional prescriptions, we show how one can deal with the matrix-elliptic operator and as a result the non-trivial computation of multiplicative anomaly can be avoided. We compute the effective potential for both massive  scalar QED theory at $d=4$ and the CW effective potential for massless case with constant background at $d=3$, and $4$. We show that the effective potential is gauge independent in all these cases, as expected.  

\section{Functional Evaluation of One Loop Effective Action}\label{sec:JD-pot}
We evaluate the one loop effective potential for the massless scalar QED case in 
Lorentz gauge where the ghosts decouple following the proposal in \cite{Dolan:1974gu,Jackiw:1974cv}. Here, the gauge parameter $\xi$ appears in the gauge fixing term, and latter show up in the effective potential through the coupling between gauge bosons and scalars. The working Lagrangian for the massless scalar QED theory in $R_\xi$ gauge\cite{Fujikawa:1972fe} is \cite{Dolan:1974gu,Jackiw:1974cv}
\begin{equation}
	\mathcal{L} = \frac{1}{2}\partial_{\mu}\phi_{i}\partial^{\mu}\phi_{i} - \frac{1}{4}F_{\mu\nu}F^{\mu\nu} - \frac{\lambda}{4!}\phi^4 - e\epsilon_{ij}\partial_{\mu}\phi_{i}\phi_{j}A^{\mu} + \frac{1}{2}e^2\phi^2A^2 - \frac{1}{\xi}(\partial_{\mu}A^{\mu})^2,
\end{equation}
where $\phi^2 = \phi_{1}^{2}+\phi_{2}^{2}, \phi^4 = (\phi_{1}^{2}+\phi_{2}^{2})^{2}$. Here $\phi_{i} (i = 1,2)$ are two real scalar fields.\\
Now we shift the scalar field by a constant background value as $\phi_{i}\rightarrow\eta_{i} + \hat{\phi}_i$. 
The effective potential $V_{1L}^{(4)} (\hat \phi)$ obtained by integrating out the fluctuations associated with $(\phi, A)$
is 
\bea
V_{1L}^{(4)} (\hat \phi) &=&- \frac{i}{2} \int\frac{d^{4}k}{(2\pi)^{4}} {\rm log [Det}[ i{\cal D}_{0}^{-1}]],
\eea
where
\begin{equation}
	i{\cal D}_{0}^{-1}[\hat{\phi},k] = \begin{pmatrix}  (k^{2} -\frac{1}{6}\lambda\hat{\phi}^{2})\delta_{ij} - \frac{1}{3}\lambda\hat{\phi}_i\hat{\phi}_j & ie\epsilon_{ii'}\hat{\phi}_{i'}k_{\nu} \\ -ie\epsilon_{jj'}\hat{\phi}_{j'}k_{\mu} & (-k^{2} + e^{2}\hat{\phi}^{2})g_{\mu\nu}+(1-\frac{1}{\xi})k_{\mu}k_{\nu}\end{pmatrix}.
\end{equation}
We rewrite $ i{\cal D}_{0}^{-1}$ in the block-diagonal form in $\{i,j\}$ and $\{\mu,\nu\}$ space employing suitable similarity transformation\footnote{Our primary expectation is that this diagonalization will not affect the computation of effective Lagrangian as it is allowed within ``Det" operator. We will come back to this point in subsequent section and discuss its possible impact.} 
\begin{equation}
	i{\cal D}_{0}^{-1} = \begin{pmatrix} iD^{-1}_{ij}(\hat{\phi},k) + iQ_{ij}(\hat{\phi},k) & 0 \\ 0 & i\Delta^{-1}_{\mu\nu}(\hat{\phi},k)    
	\end{pmatrix},
\end{equation}
where
\begin{align}
	Q_{ij}(\hat{\phi},k) &= M^{\mu}_{i}(\hat{\phi},k)\Delta_{\mu\nu}(\hat{\phi},k) M^{\nu}_{j}(\hat{\phi},-k), \nonumber \\
	iD^{-1}_{ij}(\hat{\phi},k) &= (k^{2} -\frac{1}{6}\lambda\hat{\phi}^{2})\delta_{ij} - \frac{1}{3}\lambda \hat{\phi}_{i} \hat{\phi}_{j}, \nonumber \\
	i\Delta^{-1}_{\mu\nu}(\hat{\phi},k) &= (-k^{2} + e^{2}\hat{\phi}^{2})g_{\mu\nu}+(1-\frac{1}{\xi})k_{\mu}k_{\nu}, \nonumber \\
	M^{\mu}_{i}(\hat{\phi},k) &= ie\epsilon_{ii'}\hat{\phi}_{i'}k^{\mu}.
	\label{eq5}
\end{align}

The one loop contribution to effective potential comes from the following expression
\bea 
	V_{1L}^{(4)} (\hat \phi) &=& -\frac{i}{2}\int\frac{d^{4}k}{(2\pi)^{4}} \text{log [Det}\, [i{\cal D}_{0}^{-1}]] \nn\\
	& =&-\frac{i}{2}\int\frac{d^{4}k}{(2\pi)^{4}} \text{log Det}[i\Delta^{-1}_{\mu\nu}(\hat{\phi},k) \cdot (iD^{-1}_{ij}(\hat{\phi},k) + iQ_{ij}(\hat{\phi},k))].
	\label{eq6}
\eea
We factorise the determinant in the following form, equivalent to prescription provided in \cite{Jackiw:1974cv, Dolan:1974gu}
\be \label{MPA-1}
\text{ Det}[i\Delta^{-1}_{\mu\nu}(\hat{\phi},k) \cdot (iD^{-1}_{ij}(\hat{\phi},k) + i Q_{ij}(\hat{\phi},k))]=\text{ Det}[i\Delta^{-1}_{\mu\nu}(\hat{\phi},k)]] \text{ Det}[ (iD^{-1}_{ij}(\hat{\phi},k) + i Q_{ij}(\hat{\phi},k))].
\ee
We  feed \eqref{MPA-1} in  \eqref{eq6}, and then along with \eqref{eq5} we find
\begin{equation}\label{Vk4}
	V_{1L}^{(4)} (\hat \phi) = -\frac{i}{2}\int\frac{d^4k}{(2\pi)^{4}} \,\text{log}\,[(k^{2} - \mathcal{A}_{1}^{2})(k^{2} - \mathcal{A}_{2}^{2})(k^{2} - \mathcal{A}_{3}^{2})(k^{2} - \mathcal{A}_{4}^{2})^3],
\end{equation}
where
\begin{eqnarray}
	\mathcal{A}_{1}^{2} &=& \frac{1}{12}(\lambda\hat{\phi}^{2} + \hat{\phi}^{2}\sqrt{\lambda^{2} - 24\xi\lambda e^{2}})\,,\quad
	\mathcal{A}_{3}^{2} = \frac{1}{2}\lambda\hat{\phi}^{2}\,,\quad \nn \\
	\mathcal{A}_{2}^{2} &=& \frac{1}{12}(\lambda\hat{\phi}^{2} - \hat{\phi}^{2}\sqrt{\lambda^{2} - 24\xi\lambda e^{2}})\,,\quad 
\mathcal{A}_{4}^{2} = e^{2}\hat{\phi}^{2}\,.
\end{eqnarray}

After regularising the integral and employing the $\overline{MS}$ renormalization, we find the effective potential as

\begin{equation}\label{eq:veff-dj}
	V_{1L}^{(4)}(\hat{\phi})\Big |_{DJ} = \frac{\hat{\phi}^{4}}{4!}\Big[\frac{1}{8\pi^{2}}\Big(\frac{5}{6}\lambda^{2} + 9e^{2} - \xi \lambda e^{2} \Big)\,\text{log} \Big(\frac{\hat{\phi}^{2}}{\mu^2} \Big)\Big].
\end{equation}
We see that the effective potential is dependent on the gauge parameter $\xi$ as first pointed out in \cite{Jackiw:1974cv}. In the next section we show that the gauge dependent term arises due to the assumption that the determinants can be factored as in \eqref{MPA-1} does not hold due to regualrization. The correction term is called the multiplicative anomaly, and for the Abelian Higgs example that we are considering the extra multiplicative anomaly term will precisely cancel the $\xi$ dependent term in   \eqref{eq:veff-dj} to give us an gauge independent one loop potential.

\section{Zeta-function Regularization and Multiplicative Anomaly}\label{sec:zeta-function-ma}

Let us define an elliptic operator $\frac{\delta^2 \mathcal{L}}{\delta \Phi^2}\equiv \Delta \equiv D^2+M^2+U$ in d-dimensional  Euclidean space. Then the one loop effective action is proportional to $\text{\text{log[Det}}[\Delta]]$ which is also, in general, equivalently written as $\text{Tr[log}[\Delta]]$. In case of a theory with multiple dynamical quantum fields, the effective action is reduced to the following form 
\begin{eqnarray}
\text{log[Det}[\Delta_1\; \Delta_2\; \Delta_3 \; \cdots]] \equiv \text{Tr[log}[\Delta_1\; \Delta_2\; \Delta_3 \; \cdots]],
\end{eqnarray}
where each of the $\Delta_i$'s are elliptic operator of order two. To compute this effective action we employ the following relation 
\begin{eqnarray}\label{det}
\text{log[Det}[\Delta_1\; \Delta_2\; \Delta_3 \; \cdots]] = \text{log[Det}[\Delta_1] + \text{log[Det}[\Delta_2]] + \cdots.
\end{eqnarray}
The above relation in turn implies that 
\begin{eqnarray}\label{eq:trace}
\text{Tr[log[}\Delta_1\; \Delta_2\; \Delta_3 \; \cdots]] = \text{Tr[log[}\Delta_1]] + \text{Tr[log[}\Delta_2] + \cdots.
\end{eqnarray}
The above two equations (\ref{eq:trace}), and (\ref{det}) are not true in general, as $\text{log}[\Delta_i]$ is not a trace-class operator and  the trace, here,  is the zeta-regularised trace ($\text{Tr}_\zeta$) which is not linear unlike the ordinary traces \cite{Weinberg:1973ua}. 
This can be understood through a simple well-known example with the following matrices \cite{Bytsenko:1994bc}
\begin{eqnarray}
	A &=& diag\{1,2,3,\cdots\},\quad B = diag\{1,1,1,\cdots\}, \nonumber \\
		C &=& A + B = diag\{2,3,4,\cdots\},
\end{eqnarray}
Now, if we compute the regularised zeta-trace of these matrices we find
\begin{eqnarray}
	Tr_\zeta(A)= \zeta(-1)&=& -\frac{1}{12}, \quad Tr_\zeta(B)= \zeta(0)= -\frac{1}{2}, \nn\\
		Tr_\zeta(A+B)=	Tr_\zeta(C)& =& \zeta(-1)-1= -\frac{13}{12} \neq Tr_\zeta(A) + Tr_\zeta(B). 
\end{eqnarray}
Thus, whenever we decompose the $\text{log[Det}[\cdots]]$ of a product of elliptic operators into $\text{log[Det}[\cdots]]$ of individual ones, e.g., $\text{log[Det}[\Delta_1 \; \Delta_2]]\equiv \text{log[Det}[\Delta_1]]+\text{log[Det}[\Delta_2]]$, we must introduce a compensating factor, known as {\it Multiplicative Anomaly} $\mathbb{A}[\Delta_1, \Delta_2]$ as follows
\begin{eqnarray}\label{eq:logdet}
\text{log[Det}[\Delta_1 \; \Delta_2]] =  \text{log[Det}[\Delta_1]]+\text{log[Det}[\Delta_2]] + \mathbb{A}[\Delta_1, \Delta_2]. 
\end{eqnarray}
The  multiplicative anomaly  $\mathbb{A}[\Delta_1, \Delta_2]$ , for compact as well as non-compact manifold can be computed with  $\zeta-$function regularisation \cite{Hawking:1976ja,Elizalde-2,Elizalde-3} and Wodzicki residue formula \cite{Wodzicki,Mickelsson:1994fb,Bytsenko:1994bc,Elizalde:1997nd}. 
Employing zeta-function regularisation, the one loop effective Lagrangian is given as  \cite{Bytsenko:1994bc}
\begin{eqnarray}
	\mathcal{L}_{eff}^{1L} &=& c_s\; \text{log[Det}[\Delta/\mu^2]] = c_s \Big[\zeta^{'}(0)+\zeta(0) \text{log}(\mu^2)\Big],
\end{eqnarray}
where $\mu$ is the renormalisation scale and $\Delta$ is the strong elliptic operator. In terms of regularised zeta-function one can recast the multiplicative anomaly as \cite{Bytsenko:1994bc}
\begin{eqnarray}
\mathbb{A}[\Delta_1, \Delta_2] &=& \frac{d}{dt} \Big[\zeta(t|\Delta_1 \Delta_2)-\zeta(t|\Delta_1)-\zeta(t|\Delta_2)\Big]\Big |_{t=0},
\end{eqnarray}
where $t$ is the Schwinger parameter\footnote{We use the same Schwinger parameter $t$ in latter section where we  discuss Heat-Kernel based method to compute the effective action.} associated with the heat equation
\begin{equation}\label{eq:heat_eq}
	\left(\partial_t+\Delta_x\right)K(t,x,y,\Delta)=0, \quad  K(0,x,y,\Delta)=\delta(x-y),
\end{equation}
with $K(t,x,y,\Delta)$ is the Heat-kernel and $\Delta$ is the (strong) elliptic operator .
The $\zeta$-function of an elliptic operator is related to its Heat-Kernel coefficients (HKCs) \cite{Belkov:1995gjw,Vassilevich:2003xt,Avramidi:2001ns} and the multiplicative anomaly can be written in terms of the HKCs \cite{Seeley}.  Thus,  It is evident that in case of non-compact manifold $\mathbb{A}[\Delta_1, \Delta_2]=0$ when either $d=2$ and(or) odd integer due to lack of singularity of the zeta-function at $t=0$ in both cases. One can also estimate this anomaly following the footsteps of Ref.~\cite{Kontsevich:1994xe}
\begin{eqnarray}
\mathbb{A}[\Delta_1, \Delta_2] &=& \frac{r_2}{2r_1 (r_1+r_2)} Res\Big[\Big(\text{log}(\Delta_1 \Delta_2^{-r_1/r_2})\Big)^2\Big],
\end{eqnarray}
where $\Delta_i \equiv (k^2+\mathcal{A}_i^2)$ are two elliptic operators of order $r_1$ and $r_2$ respectively, and the residue($Res$) is computed in the large $k$-limit. This can be further expressed as the following form 
employing the Wodzicki residue \cite{Wodzicki} formula as
\begin{eqnarray}
\mathbb{A}[\Delta_1, \Delta_2] &=& \frac{1}{2r_1 r_2 (r_1+r_2)} Res\Big[\Big(\text{log}(\Delta_1^{r_2} \Delta_2^{-r_1})\Big)^2\Big].
\end{eqnarray}
One can express the multiplicative anomaly factor in a simplified form  in the large $k$-limit as \cite{Bytsenko:1994bc}
\begin{eqnarray}\label{eq:muti-anomaly}
	\mathbb{A}[\Delta_1, \Delta_2] \supset \frac{(-1)^q\; \mathcal{V}_d}{4 (4\pi)^q \Gamma(q)} \sum_{n=1}^{q-1} \frac{1}{n(q-n)} [(\mathcal{A}_1^2)^n-(\mathcal{A}_2^2)^n] [(\mathcal{A}_1^2)^{q-n}-(\mathcal{A}_2^2)^{q-n}],
\end{eqnarray}
where, $q=d/2>1$, and $\mathcal{V}_d$ is the Euclidean  $d$-dimensional volume.

\subsection{Effective Action with Multiplicative Anomaly}

We carefully investigate the method adopted by Dolan-Jackiw  \cite{Jackiw:1974cv, Dolan:1974gu} to compute the effective action suffers similar issue. First they have diagonalised the matrix-elliptic operator $\Delta=\Delta_1 \Delta_2\dots$ and then rewritten the $\text{log[Det}[ \Delta]]$ of the matrix-elliptic operator as a sum of $\text{log[Det}[\Delta_i]]$ of individual elliptic operators. As we have discussed earlier, this decomposition must be compensated by the multiplicative anomaly. We propose that the gauge-parameter dependence  of the one loop effective potential that computed by Dolan-Jackiw, see Sec.\ref{sec:JD-pot}, is a reflection of this missing part that should be added through the multiplicative-anomaly. The operator in the Jackiw-Dolan effective theory \eqref{eq6} is 
\be
\text{log Det}[\, i{\cal D}_{0}^{-1}(\hat{\phi},k)]=\text{log Det}[\Delta_1 \; \Delta_2 \; \Delta_3\; \Delta_4],
\ee
where 
\bea
\Delta_1  = k^2 +\frac{1}{12} \hat{\phi}^2[\lambda+\sqrt{\lambda^2-24\xi \lambda e^2}]&=& k^2+ \mathcal{A}_1^2\,,  \quad \Delta_3  = k^2 + \frac{1}{2} \lambda \hat{\phi}^2, \nn\\
\Delta_2  = k^2 + \frac{1}{12} \hat{\phi}^2[\lambda-\sqrt{\lambda^2-24\xi \lambda e^2}] &=& k^2 + \mathcal{A}_2^2 \,, \quad 
\Delta_4  = (k^2 + e^2\hat{\phi}^2)^3.
\eea

Following  Ref.~\cite{Bytsenko:1994bc}, the effective potential requires the computation of 
\begin{eqnarray}
\text{log[Det}[\Delta_1 \; \Delta_2 \; \Delta_3\; \Delta_4]] &=& \text{log[Det}[\Delta_1]] +  \;  \text{log[Det}[\Delta_2]] +\; \text{log[Det}[ \Delta_3]]+\;  \text{log[Det}[\Delta_4]], \nonumber \\
&+&\mathbb{A}[\Delta_1, \Delta_2, \Delta_3,\Delta_4],
\end{eqnarray} 
where $\mathbb{A}[\Delta_1, \Delta_2, \Delta_3,\Delta_4]$ is the multiplicative anomaly.  Thus, we primarily focus on the computation of $\mathbb{A}[\Delta_1, \Delta_2]$ to note down the gauge parameter dependent contributions to justify our previous claim. Following the prescription of \cite{Bytsenko:1994bc, Elizalde:1997nd}\footnote{For the sake of simplicity, while defining the anomaly, we ignore the presence of renormalisation scale $\mu$ that enters as $\text{log}(\hat{\phi}^2/\mu^2)$ when we employ the zeta-regularization,  very similar to the computation of other regular $\text{log[Det}[\Delta_i]]$ terms \cite{Elizalde:1998xq}.}, the $\xi$-dependent term can be extracted from
anomaly function given in (\ref{eq:muti-anomaly}). 
 In $d=4$-dimensional Euclidean space, i.e., $q=2$, we find, the $\xi$-dependent anomaly density function,  using \eqref{eq:muti-anomaly},  that contributes to the effective Lagrangian
\begin{eqnarray}
	\mathbb{a}[\Delta_1, \Delta_2] (\xi)=	\frac{\mathbb{A}[\Delta_1, \Delta_2] (\xi)}{\mathcal{V}_4} &=&  \frac{(-1)^2}{4 (4\pi)^2 \Gamma(2)}   [(\mathcal{A}_1^2)-(\mathcal{A}_2)^n] [(\mathcal{A}_1^2)-(\mathcal{A}_2^2)] \nonumber \\
&=& 	 -\frac{2\xi \; e^2\;\lambda \hat{\phi}^4}{4! (8\pi^2)}. 
\end{eqnarray}
Thus, the effective Lagrangian density takes the following functional form
\begin{eqnarray}
	\mathcal{L}_{eff}^{1L} &\equiv & c_s \text{log[Det}[\Delta_1 \Delta_2]]=  c_s\Big[ \text{log[Det}\Delta_1]] +  \text{log[Det}\Delta_2]] + 	\mathbb{a}[\Delta_1, \Delta_2] (\xi)\Big] .
\end{eqnarray}
Then, after encoding the missed-out  contributions due to zeta-regularised multiplicative-anomaly we find the one loop regularised effective potential as
\begin{eqnarray}
	V_{eff}^{1L} &=&   	V_{1L}^{(4)}(\hat{\phi})\Big |_{DJ} 	-c_s \; \mathbb{a}[\Delta_1, \Delta_2] (\xi) \text{log}(\hat{\phi}^2/\mu^2) \nn \\
	&=& 	 \frac{1}{4!} \dfrac{\hat{\phi}^4}{8\pi^2} \Bigg\{ \Bigg[ \dfrac{5 \lambda^2}{6}   + 9 e^4 -\xi \; e^2\;\lambda   \Bigg] -\frac{1}{2}\Bigg[ -\frac{2\xi \; e^2\;\lambda \hat{\phi}^4}{4! (8\pi^2)}\Bigg]\Bigg\} \text{log}(\hat{\phi}^2/\mu^2) \nn \\
 &=&	 \frac{1}{8} \dfrac{\hat{\phi}^4}{8\pi^2}  \Bigg[ \dfrac{5 \lambda^2}{18}   + 3 e^4 \Bigg] \text{log}(\hat{\phi}^2/\mu^2),
\end{eqnarray}
that is gauge-parameter ($\xi$) independent. Here, $	V_{1L}^{(4)}(\hat{\phi})\Big |_{DJ}$ is the effective potential computed in \cite{Jackiw:1974cv,Dolan:1974gu}, see \eqref{eq:veff-dj}, and $c_s=1/2$ for real scalar.
 
\subsection{Effective Action in $d = 2$ and $3$ Dimensions}
In $d=2$, we work with the following Lagrangian
\begin{equation}
	\mathcal{L} = \frac{1}{2}\partial_{\mu}\phi_{i}\partial^{\mu}\phi_{i} - \frac{1}{4}F_{\mu\nu}F^{\mu\nu} - \frac{\lambda}{4!}\phi^4 - e\epsilon_{ij}\partial_{\mu}\phi_{i}\phi_{j}A^{\mu} + \frac{1}{2}e^2\phi^2A^2 - \frac{1}{\xi}(\partial_{\mu}A^{\mu})^2.
\end{equation}
Here, $\phi$ and $A_\mu$ have no mass dimension, but the couplings $\lambda$, and $e$ has mass dimensions 2 and 1 respectively. 

Now we evaluate the quantum corrections corresponding to the $\phi^2$ operator and that reads as
\begin{equation}\label{Vk4}
	V_{1}(\hat{\phi}) = -\frac{i}{2}\int\frac{d^2 k}{(2\pi)^{4}} \,\text{log}\,[(k^{2} - \mathcal{A}_{1}^{2})(k^{2} - \mathcal{A}_{2}^{2})(k^{2} - \mathcal{A}_{3}^{2})(k^{2} - \mathcal{A}_{4}^{2})].
\end{equation}
Following the same prescription till now we reach at the following result 
\begin{equation}
	V^{(2)}_{1L}(\hat{\phi})|_{\hat{\phi}^2} = -\hat{\phi}^{2}\frac{1}{8\pi}(\frac{2}{3}\lambda + e^{2})\text{log}(\hat{\phi}^{2}).
\end{equation}
 It is important to note that  the term $V^{(2)}_{1L}(\hat{\phi})|_{\hat{\phi}^2}$ is gauge parameter $\xi$ independent. In two-dimension, multiplicative anomaly is zero. In case of massless scalar QED in $d=2$, where $\phi$ is dimensionless, we can compute the one loop corrections corresponding to the $\phi^{2n},\; \forall n  \in \mathbb{Z}^{+}$ operator \footnote{To do so we need to have $\frac{\lambda}{(2n+2)!} \phi^{2n+2}$ term in the Lagrangian.}, and that term will also be independent of $\xi$. In other words,  the equation (\ref{eq:logdet}) holds in two-dimension and one is not required to compensate with the multiplicative anomaly. This is consistent with our observation.  

Another important property of multiplicative anomaly is that $\mathbb{A}_{d}$ vanishes for odd dimensional non-compact manifold.  
We verify this  explicitly employing the Jackiw-Dolan functional method.
In $d=3$, we have the Lagrangian
\begin{equation}\label{eq:lag3}
	\mathcal{L} = \frac{1}{2}\partial_{\mu}\phi_{i}\partial^{\mu}\phi_{i} - \frac{1}{4}F_{\mu\nu}F^{\mu\nu} - \frac{\lambda}{6!}\phi^{6} - e\epsilon_{ij}\partial_{\mu}\phi_{i}\phi_{j}A^{\mu} + \frac{1}{2}e^2\phi^{2}A^{2} - \frac{1}{\xi}(\partial_{\mu}A^{\mu})^{2}.
\end{equation}
Here, $\phi,A_\mu$, and $e$ have  mass dimensions 1/2 each, where the coupling $\lambda$ is dimensionless. We define $\phi^{2} = \phi_{1}^{2}+\phi_{2}^{2}, \phi^{6} = (\phi_{1}^{2}+\phi_{2}^{2})^{3}$.
Following the same prescription till now we find
\begin{equation}
	V_{1}(\hat{\phi}) = -\frac{i}{2}\int\frac{d^{3}k}{(2\pi)^{3}} \text{log}[(k^{2} - \mathcal{A}_{1}^{2})(k^{2} - \mathcal{A}_{2}^{2})(k^{2} - \mathcal{A}_{3}^{2})(k^{2} - \mathcal{A}_{4}^{2})^{2}],
\end{equation}
where
\begin{align}
	\mathcal{A}_{1}^{2} &= \frac{1}{240}(\lambda\hat{\phi}^{4} + \sqrt{\lambda^{2}\hat{\phi}^{8} - 480\alpha\lambda e^{2}\hat{\phi}^{6}});
	\mathcal{A}_{2}^{2} = \frac{1}{240}(\lambda\hat{\phi}^{4} - \sqrt{\lambda^{2}\hat{\phi}^{8} - 480\alpha\lambda e^{2}\hat{\phi}^{6}}); \notag \\
	\mathcal{A}_{3}^{2} &= \frac{1}{4!}\lambda\hat{\phi}^{4};
	\mathcal{A}_{4}^{2} = e^{2}\hat{\phi}^{2}
\end{align}
The one loop effective potential associated with $\hat{\phi^{6}}$ term (marginal operator) is given as
\begin{equation}\label{eq:veff-3}
	V_{1L}^{(3)}|_{\hat{\phi}^6} = -\frac{1}{12\pi}\lambda^{\frac{3}{2}}[(\frac{1}{5!})^{\frac{3}{2}}+(\frac{1}{4!})^{\frac{3}{2}}]\hat{\phi}^{6},
\end{equation}
which is independent of $\xi$-parameter in consistent with vanishing multiplicative anomaly $\mathbb{A}_{d=3}=0$.

We have noted that computation of multiplicative in generic cases is rather difficult. Thus, instead of employing that we will discuss a new prescription to compute the one loop effective action where we will deal with the full matrix-elliptic operator, such that we can avoid computation of multiplicative-anomaly, separately.  

\section{ Effective Action using Heat-Kernel Expansion}\label{sec:HK-Veff}

 The Heat-Kernel (HK)  \cite{Seeley,Belkov:1995gjw,Vassilevich:2003xt,Avramidi:2001ns,Kontsevich:1994xe} 
 for an elliptic operator $ \Delta \equiv\frac{\delta^2 \mathcal{L}}{\delta \Phi^2}=D^2+M^2+U$,
 is the solution of the
 the heat equation and initial condition given by,
\begin{equation}\label{eq:heat_eq}
    \left(\partial_t+\Delta_x\right)K(t,x,y,\Delta)=0, \quad  K(0,x,y,\Delta)=\delta(x-y).
\end{equation}
Using the Heat-Kernel method  the effective Lagrangian is given as \cite{Avramidi:1990je, Banerjee:2023iiv,Banerjee:2023xak,Banerjee:2024rbc,Adhikary:2025pbb,Chakrabortty:2023yke}
\begin{equation}\label{Leff}
    \mathcal{L}_{eff}= c_s\;\; \text{tr}\; \int_0^{\infty} \frac{dt}{t} K(x,x,\Delta,t),
\end{equation}
where $c_s=1/2(1)$ for real(complex) bosons. The integration over $t$ which is equivalent to the  momentum integral in \eqref{eq6} is performed after first expressing $K(x,x,\Delta,t)$ as a power series in $t$,
\begin{equation}\label{eq:interaction}
    K(x,x,\Delta,t)=\sum_{k=0}^\infty \frac{(-t)^k}{k\,!}b_k(x),
\end{equation}
the $b_k(x)$ are the Heat-Kernel coefficients which are polynomials of  $U$, and $D$. In the following section we compute the Heat-Kernel coefficients of the Abelian Higgs model to obtain the effective action.

\subsection{Effective Action of the Abelian Higgs Model}
We consider an Abelian Higgs model with  an $O(2)$ scalar multiplet $\phi_a$ and the gauge field  $A_\mu$,
\bea\label{AHM}
    \mathcal{L} &=&\frac{1}{2}\partial_\mu \phi_a  \partial^{\mu}\phi_a - \frac{1}{4}F_{\mu \nu}F^{\mu \nu} - \frac{1}{2}M_1^2 \phi^2 - \frac{\lambda}{4!}\phi^4 - e \epsilon_{ab}A^\mu \phi_b  \partial_\mu \phi_a\nn\\  
    &+& \frac{1}{2} M_2^2 A^2  + \frac{1}{2} e^2 \phi^2 A^2 -\frac{1}{2 \xi}(\partial_\mu A^\mu)^2,
\eea
where we have introduced the masses $M_1$ and $M_2$ $\in \mathbb{R}^+$ as infrared regulators. The last term in (\ref{AHM}) is the gauge fixing term in the $R_\xi$ gauge where the ghost Lagrangian decouples from $\phi_a, A_\mu$ and we will drop the ghost Lagrangian for the one loop effective action calculation.
We combine the scalar and gauge fields to define the field multiplet $\Phi=(\phi_a,A_\mu)^T$
and  define the operator matrix 
\bea
    \Delta_{ij}=\dfrac{ \partial^2 \mathcal{L}}{\partial \Phi_i \Phi_j} =\begin{pmatrix}
    	(\partial^2 +  M_1^2 + \frac{\lambda}{6}\hat{\phi}^2 )\delta_{ab} + \frac{\lambda}{3}\hat{\phi}_a \hat{\phi}_b & e \epsilon_{ab^\prime} \partial_\nu\hat{\phi}_{b^\prime}\\
    	-e \epsilon_{a{^\prime}b} \partial_\mu \hat{\phi}_{a^\prime} & -(\partial^2 + e^2 \hat{\phi}^2+M_2^2) \delta_{\mu \nu}  + (1- \frac{1}{\xi})\partial_\mu \partial_\nu
    \end{pmatrix}  ,  \nn
\eea  
 where $(i,j)=\{(a,b),(\mu,\nu)\}$, $\partial^2$ is the 4-dimensional Laplacian in the Euclidean space $R^4$. We define 
\bea\label{M2U-1}
    M^2 = \begin{pmatrix}
       M_1^2\,\, \delta_{ab} 
       & 0 \\
       0 & -M_2^2 \,\, \delta_{\mu \nu}
   \end{pmatrix}\quad  {\rm and } \quad
   U = \begin{pmatrix}
       \frac{\lambda}{6}\hat{\phi}^2 \delta_{ab} + \frac{\lambda}{3}\hat{\phi}_a \hat{\phi}_b & e \epsilon_{ab^\prime} \partial_\nu\hat{\phi}_{b^\prime}\\
        -e \epsilon_{a{^\prime}b} \partial_\mu \hat{\phi}_{a^\prime} & -e^2 \hat{\phi}^2 \delta_{\mu \nu} + (1- \frac{1}{\xi})\partial_\mu \partial_\nu
   \end{pmatrix},
\eea
to define the Heat-Kernel.
Here, in general $M^2$ matrix does not commute with the $U$-matrix. Thus, we can not separate the free and interaction part while defining the Heat-Kernel. This happens when we there are non-degenerate massive fields present in the theory we are interested in. In that scenario, we need to define the HK employing momentum space integral  representation where the trace \footnote{This trace is defined over all internal symmetries.} of HK  can be expressed as \cite{Banerjee:2023xak}
\begin{eqnarray}
    \text{tr}\,K(t,x,x,\Delta)&=& \text{tr}\, \int \frac{d^4p}{(2 \pi)^4} \langle x| e^{-M^2 t}\, \mathcal{T} \text{exp}\Big[ -  \int_0^t \big(\partial^2 + e^{M^2 t^\prime} U e^{-M^2 t^\prime} \big) dt^\p \Big] |p \rangle \langle p|x \rangle \n \\
    &=& \text{tr}\, \int \frac{d^4p}{(2 \pi)^4} e^{-M^2 t} \langle x|\, \mathcal{T} \text{exp}\Big[ - \int_0^t \big(\partial^2 + e^{M^2 t^\prime} U e^{-M^2 t^\prime} \big) dt^\p \Big] |p \rangle e^{ipx} \n \\ 
    &=& \text{tr}\, \int \frac{d^4p}{(2 \pi)^4} e^{-M^2 t} \langle x|p \rangle \, e^{ipx}\, \mathcal{T} \text{exp}\Big[ - \int_0^t \big((\partial_\mu + ip_\mu)^2 + e^{M^2 t^\prime} U e^{-M^2 t^\prime} \big) dt^\p \Big] \n \\
    &=& \text{tr}\, \int \frac{d^4p}{(2 \pi)^4} e^{-M^2 t} \, e^{p^2 t} \mathcal{T} \text{exp}\Big[ - \int_0^t \big(\partial^2 + 2ip.\partial + e^{M^2 t^\prime} U e^{-M^2 t^\prime} \big) dt^\p \Big],
\end{eqnarray}
where we have used the property $e^{-i p \hat x} \partial_\mu  e^{i p \hat x} \equiv \partial_\mu+ip_\mu$. We note that the derivative operators $\partial_\mu$ in $U$  (eqn (\ref{M2U-1}))  are also translated by $ip_\mu$. We redefine $p^2 t$ to $p^2$ to obtain,
\begin{eqnarray}
    \text{tr}\,K(t,x,x,\Delta) = \text{tr}\, \int \frac{d^4p}{(2 \pi)^4\, t^2} e^{-M^2 t} e^{p^2 } \, \mathcal{T} \text{exp}\Big[ - \int_0^t \big(\partial^2 + 2ip.\partial/\sqrt{t} + e^{M^2 t^\prime} U e^{-M^2 t^\prime} \big) dt^\p \Big]. \nn
\end{eqnarray}
Here, $p^2= \delta_{\mu \nu}p^\mu p^\nu=-\sum_{i=1}^{d} p_i^2$\footnote{We work with all negative Euclidean metric.}.
The chronologically ordered exponential is written as 
\begin{eqnarray}
    \mathcal{F}(t,A) &=& \mathcal{T} \text{exp}\big(-\int_0^t A(t^\p) dt^\p \, \big)\n \\
    &=& 1 + \sum_{n=1}^{\infty} (-1)^n f_n(t,A) ,
\end{eqnarray}
where $f_n$ can be recast in terms of an integral equation as follows 
\begin{align}
    f_n(t,A)=\int_0^t ds_1 \int_0^{s_1} ds_2\, ...\int_0^{s_{n-1}} ds_n A(s_1)A(s_2)\,... \, A(s_n) .
\end{align}

For the choice of $M^2$ and $U$ given in (\ref{M2U-1}) we find  $A(t)$ as
\begin{align}
    A =\! \begin{pmatrix}
       (\partial^2 \!+\! \frac{2ip.\partial}{\sqrt{t}} + \frac{\lambda}{6}\hat{\phi}^2)\delta_{ab} + \frac{\lambda}{3}\hat{\phi}_a \hat{\phi}_b \!\!\!\! & e \epsilon_{ab^\prime} \partial_\nu \hat{\phi}_{b^\prime} + \frac{ie} {\sqrt{t}}\epsilon_{ab^\prime} p_\nu \hat{\phi}_{b^\prime}\\
        -e \epsilon_{a{^\prime}b} \partial_\mu \hat{\phi}_{a^\prime} -\frac{ie} {\sqrt{t}}\epsilon_{ab^\prime} p_\nu \hat{\phi}_{b^\prime} & -(\partial^2 \!+\! 
        \frac{2ip.\partial}{\sqrt{t}} + e^2 \hat{\phi}^2) \delta_{\mu \nu} + (1\!-\! \frac{1}{\xi})(\partial_\mu \partial_\nu - \frac{p_\mu p_\nu}{t} + \frac{2ip_\mu \partial_\nu}{\sqrt{t}})
   \end{pmatrix}.
\end{align}
Here, $\phi_a(x)= \hat{\phi}_a(x)+\eta_a(x)$ with $\hat{\phi}_a$ is the background field, not necessarily constant in space-time, and $\eta_a$ are the fluctuations around this background. We discuss the constant background field considering the massless scalar QED scenario in the following section that mimics the Dolan-Jackiw computation of effective potential \cite{Jackiw:1974cv,Dolan:1974gu}.  We further set the background field for the gauge field $A_\mu$ to be zero without loss of generality. 
The effective action is expressed in terms of the Heat-Kernel series as  \cite{Banerjee:2023xak}
\begin{eqnarray}
    \mathcal{L}_{eff}&=&c_s \text{tr} \int_0^\infty \frac{dt}{t} K(t,x,x,\Delta) \n \\
    &=& c_s \text{tr} \int_0^\infty \frac{dt}{t^3} \int \frac{d^4p}{(2 \pi)^4} e^{-M^2 t} e^{p^2 t} \big[ 1 + \sum_n (-1)^n f_n(t,A)  \big]. 
\end{eqnarray}
The $U$ in our theory has a mass dimension two. So, in order to get the dimension four contribution to the effective potential, we need to calculate only the $f_2$ term in the expansion,
\begin{eqnarray}
    \mathcal{L}^{(4)}_{eff} &=& c_s \text{tr} \int_0^\infty \frac{dt}{t^3} \int \frac{d^4p}{(2 \pi)^4} e^{-M^2 t} e^{p^2 } f_2(t,A)\n \\
    &=& c_s \text{tr} \int_0^\infty \frac{dt}{t^3} \int d\Omega \int_0^\infty \frac{p^3 dp}{(2 \pi)^4} e^{-M^2 t} e^{p^2 } f_2(t,A)  \n\\
    &=& c_s \text{tr} \int_0^\infty \frac{dt}{t^3} \frac{\pi^2}{2} \int_0^\infty \frac{p^3 dp}{(2 \pi)^4} e^{-M^2 t} e^{p^2 } f_2(t,A), 
\end{eqnarray}
where $c_s =\frac{1}{2}(1)$ for real(complex) scalars. After performing the momentum and $t$ integral, we get 
\begin{eqnarray}
    \mathcal{L}^{(4)}_{eff} &=& - \frac{c_s}{(32 \pi^2)} \text{tr}\left[ U_{11}^2 \text{log}(M_1^2) + U_{22}^2 \text{log}(M_2)^2 + U_{12} U_{21} \left(1 - \dfrac{M_1^2 \text{log}M_1^2 - M_2^2\text{log}M_2^2}{M_1^2 - M_2^2}\right) \right] \n \\
    &=& - \dfrac{1}{4!} \dfrac{1}{8 \pi^2} \Bigg[ \dfrac{5\lambda^2}{6} \hat{\phi}^4 \text{log}(M_1^2)  + 9 e^4 \hat{\phi}^4  \text{log}(M_2^2) +\Big(3(1-\frac{1}{\xi})^2 \partial^4 - 6 (1- \frac{1}{\xi})e^2 \partial^2 \hat{\phi}^2 \Big)\text{log}(M_2^2) \n \\
    &+& 6 e^2 \partial_\mu \hat{\phi}_a \partial^{\mu}\hat{\phi}_a \Big(1 - \dfrac{M_1^2 \text{log}M_1^2 - M_2^2\text{log}M_2^2}{M_1^2 - M_2^2}\Big) \Bigg]. 
\end{eqnarray}
Here, we find that two operators $\partial^4$ and $\partial^2 \hat{\phi}^2$ are total derivatives \footnote{Note that in this computation $M_{1,2}^2$ are constant in space-time.}, thus do not contribute in the effective potential. The final form of the effective Lagrangian reads as
\begin{eqnarray}
	\mathcal{L}^{(4)}_{eff} 	&=& - \dfrac{1}{4!} \dfrac{1}{8 \pi^2} \Bigg[ \dfrac{5\lambda^2}{6} \hat{\phi}^4 \text{log}(M_1^2)  + 9 e^4 \hat{\phi}^4  \text{log}(M_2^2)  \n \\
	&+& 6 e^2 \partial_\mu \hat{\phi}_a \partial^{\mu}\hat{\phi}_a \Big(1 - \dfrac{M_1^2 \text{log}M_1^2 - M_2^2\text{log}M_2^2}{M_1^2 - M_2^2}\Big) \Bigg].
\end{eqnarray}
Here, the last term offers the additional contribution to field renormalization factor $Z_{\phi}$ in presence of scalar-gauge field interaction, and thus proportional to $e^2$.

\subsection{Effective Action of Scalar QED in $d=3$}
We compute the effective action in  $d=3$ dimension for the similar scalar QED theory using the Heat-Kernel method for comparison with \eqref{eq:veff-3}. The Lagrangian is (\ref{eq:lag3})
\begin{align}
	\mathcal{L}=\frac{1}{2}\partial_\mu \phi_a  \partial^{\mu}\phi_a - \frac{1}{4}F_{\mu \nu}F^{\mu \nu}  - \frac{\lambda}{6!}\phi^6 - e \epsilon_{ab}A^\mu \phi_b  \partial_\mu \phi_a + \frac{1}{2} e^2 \phi^2 A^2 -\frac{1}{2 \xi}(\partial_\mu A^\mu)^2.
\end{align}
The elliptic operator is defined as 
\bea
\Delta_{ij}=\dfrac{ \partial^2 \mathcal{L}_{_E}}{\partial \Phi_i \Phi_j} = \begin{pmatrix}
	(\partial^2 + \frac{\lambda}{5!}\hat{\phi}^4 )\delta_{ab} + \frac{\lambda}{30} \hat{\phi}^2\hat{\phi}_a \hat{\phi}_b & e \epsilon_{ab^\prime} \partial_\nu\hat{\phi}_{b^\prime}\\
	-e \epsilon_{a{^\prime}b} \partial_\mu \hat{\phi}_{a^\prime} & -(\partial^2 + e^2 \hat{\phi}^2) \delta_{\mu \nu}  + (1- \frac{1}{\xi})\partial_\mu \partial_\nu
\end{pmatrix} , 
\eea  
where $\quad \quad (i,j)=\{(a,b),(\mu,\nu)\},$, and we define 
\bea\label{M2U-3}
 M^2 = \begin{pmatrix}
	\frac{\lambda}{4!}\hat{\phi}^4\delta_{ab} & 0 \\
	0 & -e^2 \hat{\phi}^2 \delta_{\mu \nu}
\end{pmatrix} \quad  {\rm and } \quad
  U = \begin{pmatrix}
	\frac{\lambda}{30} \hat{\phi}^2\hat{\phi}_a \hat{\phi}_b - \frac{\lambda}{30}\hat{\phi}^4 \delta_{ab} & e \epsilon_{ab^\prime} \partial_\nu \hat{\phi}_{b^\prime}\\
	-e \epsilon_{a{^\prime}b} \partial_\mu \hat{\phi}_{a^\prime} & (1- \frac{1}{\xi})\partial_\mu \partial_\nu \end{pmatrix}.
  \eea
  
Employing the  HK method as discussed in the previous section, we find the effective action corresponding to the $\hat{\phi^{6}}$ term as 
\begin{eqnarray}
	\mathcal{L}^{(3)}_{eff} &=& \frac{1}{2 \pi^2} \text{tr} \Bigg( \frac{1}{3} (M_1^2)^{\frac{3}{2}} - \frac{1}{2} \sqrt{M_1^2} \, U_{11} + \frac{1}{8} \frac{U_{11}^2}{\sqrt{M_1^2}} \Bigg) \\
	&=&\dfrac{\lambda^{\frac{3}{2}}}{2 \pi} \hat{\phi}^6 \frac{1}{(4!)^{\frac{3}{2}}} \frac{86}{75}\; .
\end{eqnarray}
This explicitly shows that the effective potential is gauge parameter independent as in (\ref{eq:veff-3}), computed employing Dolan-Jackiw method. This is expected  as in $d=3$, i.e., odd space-time dimension the multiplicative anomaly vanishes.

\subsection{Gauge Invariant Coleman-Weinberg Potential }
We assume the similar Scalar QED Lagrangian in $d=4$ as adopted in \cite{Jackiw:1974cv,Dolan:1974gu}
\begin{align}
	\mathcal{L}=\frac{1}{2}\partial_\mu \phi_a  \partial^{\mu}\phi_a - \frac{1}{4}F_{\mu \nu}F^{\mu \nu}  - \frac{\lambda}{4!}\phi^4 - e \epsilon_{ab}A^\mu \phi_b  \partial_\mu \phi_a + \frac{1}{2} e^2 \phi^2 A^2 -\frac{1}{2 \xi}(\partial_\mu A^\mu)^2.
\end{align}

We now have the operator matrix 
\bea
    \Delta_{ij}=\dfrac{ \partial^2 \mathcal{L}_{_E}}{\partial \Phi_i \Phi_j} =  \begin{pmatrix}
    (\partial^2 + \frac{\lambda}{6}\hat{\phi}^2 )\delta_{ab} + \frac{\lambda}{3}\hat{\phi}_a \hat{\phi}_b & e \epsilon_{ab^\prime} \partial_\nu\hat{\phi}_{b^\prime}\\
    -e \epsilon_{a{^\prime}b} \partial_\mu \hat{\phi}_{a^\prime} & -(\partial^2 + e^2 \hat{\phi}^2) \delta_{\mu \nu} + (1- \frac{1}{\xi})\partial_\mu \partial_\nu
    \end{pmatrix}   , 
\eea  
where $(i,j)=\{(a,b),(\mu,\nu)\}$, and we choose the diagonal terms as $M^2$ and the remaining field terms in $U$ as
\bea\label{M2U-2}
  M^2 = \begin{pmatrix}
		\frac{\lambda}{2}\hat{\phi}^2\delta_{ab} & 0 \\
		0 & -e^2 \hat{\phi}^2 \delta_{\mu \nu}
	\end{pmatrix}  \quad  {\rm and } \quad
  U = \begin{pmatrix}
		\frac{\lambda}{3}\hat{\phi}_a \hat{\phi}_b - \frac{\lambda}{3}\hat{\phi}^2 \delta_{ab} & e \epsilon_{ab^\prime} \partial_\nu \hat{\phi}_{b^\prime}\\
		-e \epsilon_{a{^\prime}b} \partial_\mu \hat{\phi}_{a^\prime} & (1- \frac{1}{\xi})\partial_\mu \partial_\nu
	\end{pmatrix}.
  \eea
The for the Coleman-Weinberg potential, we take $\hat \phi$=constant and  in this case the terms of the $M^2$ matrix act as the infrared regulators.  
which gives us $A(t)$ as
\begin{align}
	A =\! \begin{pmatrix}
		(\partial^2 \!+\! \frac{2ip.\partial}{\sqrt{t}} - \frac{\lambda}{3}\hat{\phi}^2)\delta_{ab} + \frac{\lambda}{3}\hat{\phi}_a \hat{\phi}_b \!\!\!\! & e \epsilon_{ab^\prime} \partial_\nu \hat{\phi}_{b^\prime} + \frac{ie \epsilon_{ab^\prime} p_\nu \hat{\phi}_{b^\prime}}{\sqrt{t}}\\
		-e \epsilon_{a{^\prime}b} \partial_\mu \hat{\phi}_{a^\prime}  - \frac{ie \epsilon_{a{^\prime}b} p_\mu \hat{\phi}_{a^\prime}}{\sqrt{t}} & -(\partial^2 \!+\! \frac{2ip.\partial}{\sqrt{t}} ) \delta_{\mu \nu} + (1\!-\! \frac{1}{\xi})(\partial_\mu \partial_\nu - \frac{p_\mu p_\nu}{t} + \frac{2ip_\mu \partial_\nu}{\sqrt{t}})
	\end{pmatrix}.
\end{align}
Using the Heat-Kernel method now we compute the effective Lagrangian up to dimension four as
\begin{eqnarray}
	\mathcal{L}^{(4)}_{eff} &=&- \frac{1}{(64 \pi^2)} \text{tr}\Big[ M_1^4 \Big( \text{log}M_1^2 -\frac{3}{2} \Big) + M_2^4 \Big( \text{log}M_2^2 -\frac{5}{6} \Big) + 2M_1^2 U_{11} \Big( \text{log}M_1^2 -1 \Big) \nn\\
	&+& 2M_2^2 U_{22} \Big( \text{log}M_1^2 -1 \Big) +U_{11}^2 \text{l}(M_1^2) + U_{22}^2 \text{log}(M_2)^2 \n \\
	&-& U_{12} U_{21} \left(1 - \dfrac{M_1^2 \text{log}M_1^2 - M_2^2\text{log}M_2^2}{M_1^2 - M_2^2}\right) \Big] \n \\[0.3\baselineskip]
	&=& - \frac{1}{8} \dfrac{1}{8\pi^2} \Bigg[ \dfrac{5 \lambda^2}{18} \hat{\phi}^4 \Big(\text{log}\big(\frac{\lambda \hat{\phi}^2}{2}\big) -\frac{3}{2}\Big)  + 3 e^4 \hat{\phi}^4 \Big(\text{log}(e^2 \hat{\phi}^2) -\frac{5}{6}\Big) + \big(1- \frac{1}{\xi} \big)^2 \text{log}(e^2 \hat{\phi}^2) \partial^4  \nn \\
	&-& 2 \big(1- \frac{1}{\xi}\big)e^2 \Big(\text{log}(e^2 \hat{\phi}^2) -1 \Big) \partial^2 \hat{\phi}^2 + 2 e^2 \partial_\mu \hat{\phi}_a  \partial^{\mu}\hat{\phi}_a \left(1 - \dfrac{\frac{\lambda \hat{\phi}^2}{2} \text{log}(\frac{\lambda \hat{\phi}^2}{2}) - e^2\hat{\phi}^2 \text{log}e^2 \hat{\phi}^2}{\frac{\lambda \hat{\phi}^2}{2} - e^2 \hat{\phi}^2}\right)\Bigg].  \nn \\
\eea 
Since the background field is a constant in space-time, the $\partial$ terms are zero, and we are left with the Coleman-Weinberg potential for the Abelian Higgs model given by
\bea\label{VCW}
V_{CW}^{(1L)}	&=&  \dfrac{1}{4!} \dfrac{1}{8 \pi^2} \Bigg[ \dfrac{5 \lambda^2}{6} \hat{\phi}^4 \Big(\text{log}\big(\frac{\lambda \hat{\phi}^2}{2}\big) -\frac{3}{2}\Big)  + 9 e^4 \hat{\phi}^4 \Big(\text{log}(e^2 \hat{\phi}^2) -\frac{5}{6}\Big)  \Bigg].
\end{eqnarray}
We see that the Coleman-Weinberg potential for the Abelian Higgs theory does not depend upon the gauge parameter $\xi$ unlike the result of
\cite{Jackiw:1974cv,Dolan:1974gu}. Here we have expressed the CW potential in terms of the $O(2)$ multiplet $\hat \phi =( \hat{\phi}_1,  \hat{\phi}_2)^T$. We can also write this in terms of a complex field $\Phi=\frac{1}{\sqrt{2}}(\phi_1 + i\phi_2)$ and  $\Phi^*=\frac{1}{\sqrt{2}}(\phi_1 - i\phi_2)$. To make identification with the field in the broken symmetric phase we can chose $\phi_1=h$ the Higgs field, and $\phi_2=G$ the Goldstone field whose background value is zero. In this case we can write the Coleman-Weinberg potential for the Higgs field by taking $\hat \phi=(h,0)$ and substituting in \eqref{VCW}.

\section{Conclusions} In this paper, we examine the long-standing issue of the gauge dependence of the effective potential in an Abelian Higgs theory. We show that the standard path integral method calculation of the effective potential by integrating out the fluctuations over a constant background results in the computation of  the determinant of an operator quartic in $\partial^2$. In the case of the functional method, one is required to express that operator  as a product of four quadratic operators, such that the problem reduces to the computation of determinant of operators that are linear in $\partial^2$. This decomposition does not always hold as it is protected by the regularised zeta-trace ($\text{Tr}_\zeta$), which is not linear. This results in the emergence of unavoidable multiplicative anomaly, which is also related to the presence of double pole in the effective action in functional method \cite{Barvinsky:2024irk}. We correct this factorization error by including the multiplicative anomaly term and show that this extra term, which was omitted in the earlier treatments, precisely cancels the problematic gauge-dependent term in the scalar potential. As a consistency check of this idea, we compute the effective potential in $d=2$ and $3$ spacetime dimensions, where the multiplicative anomaly is zero, and we find that the effective potential obtained after a factorization of the determinant is gauge-independent. 

We also compute the effective potential using the Heat-Kernel expansion procedure, and we find that the quantum correction to the $\lambda \phi^4$ operator has $\xi$ dependent terms, which are a total divergence. In the case of the Heat-Kernel method, we compute the effective potential involving the full matrix-elliptic operator $\Delta$. Thus, unlike the functional method, we can avoid writing the effective action as a product  $\Pi_{i}\text{log[Det}[\Delta_i]]$ such that $\Delta=\Pi_{i} \Delta_i$.   As the Heat-Kernel method avoids the factorization of the functional determinant, the multiplicative anomaly appears to be total derivative \cite{Barvinsky:2024irk}. Thus, HK prescription automatically gives a gauge invariant effective potential. We also compute the Coleman-Weinberg potential in the Heat-Kernel method and find that it is gauge-independent. 

We believe that in the even-dimensional space-time non-compact manifold, the gauge dependence in the effective potential, if any,  can be rectified by employing either HK method or adding multiplicative anomaly.  We leave the proof of this statement for our future work. This result is significant as we have a procedure for computation of the effective potential of scalars at finite temperature \cite{Chakrabortty:2024wto}, which can be used for applications phase transitions without the gauge ambiguity. It is worth mentioning that our conclusion is equally applicable for non-Abelian gauge theory as well, and thus it will be important to note the impact in the context of Standard Model Higgs effective potential which we leave for the future endeavour.

\section*{Acknowledgements}
JC thanks Sabyasachi Chakraborty, Diptarka Das, Christoph Englert, Apratim Kaviraj, Kaanapuli Ramkumar, and Nilay Kundu for useful discussions. JC acknowledges support from CRG project, SERB, India. SM thanks the IIT-Kanpur grant of a Distinguished Visiting
Professor position, for support in carrying out this work.

\end{document}